\documentclass[12pt,epsf]{article}
\usepackage{amssymb,amsmath}
\usepackage{graphicx}
\begin{document}

\def\a{\alpha}
\def\b{\beta}
\def\c{\varepsilon}
\def\d{\delta}
\def\e{\epsilon}
\def\f{\phi}
\def\g{\gamma}
\def\h{\theta}
\def\k{\kappa}
\def\l{\lambda}
\def\m{\mu}
\def\n{\nu}
\def\p{\psi}
\def\q{\partial}
\def\r{\rho}
\def\s{\sigma}
\def\t{\tau}
\def\u{\upsilon}
\def\v{\varphi}
\def\w{\omega}
\def\x{\xi}
\def\y{\eta}
\def\z{\zeta}
\def\D{\Delta}
\def\G{\Gamma}
\def\H{\Theta}
\def\L{\Lambda}
\def\F{\Phi}
\def\P{\Psi}
\def\S{\Sigma}

\def\o{\over}
\def\beq{\begin{eqnarray}}
\def\eeq{\end{eqnarray}}
\newcommand{\gsim}{ \mathop{}_{\textstyle \sim}^{\textstyle >} }
\newcommand{\lsim}{ \mathop{}_{\textstyle \sim}^{\textstyle <} }
\newcommand{\vev}[1]{ \left\langle {#1} \right\rangle }
\newcommand{\bra}[1]{ \langle {#1} | }
\newcommand{\ket}[1]{ | {#1} \rangle }
\newcommand{\EV}{ {\rm eV} }
\newcommand{\KEV}{ {\rm keV} }
\newcommand{\MEV}{ {\rm MeV} }
\newcommand{\GEV}{ {\rm GeV} }
\newcommand{\TEV}{ {\rm TeV} }
\def\diag{\mathop{\rm diag}\nolimits}
\def\Spin{\mathop{\rm Spin}}
\def\SO{\mathop{\rm SO}}
\def\O{\mathop{\rm O}}
\def\SU{\mathop{\rm SU}}
\def\U{\mathop{\rm U}}
\def\Sp{\mathop{\rm Sp}}
\def\SL{\mathop{\rm SL}}
\def\tr{\mathop{\rm tr}}

\def\IJMP{Int.~J.~Mod.~Phys. }
\def\MPL{Mod.~Phys.~Lett. }
\def\NP{Nucl.~Phys. }
\def\PL{Phys.~Lett. }
\def\PR{Phys.~Rev. }
\def\PRL{Phys.~Rev.~Lett. }
\def\PTP{Prog.~Theor.~Phys. }
\def\ZP{Z.~Phys. }


\baselineskip 0.7cm

\begin{titlepage}

\begin{flushright}
ICRR-REPORT-532\\
IPMU 08-0081\\
\end{flushright}

\vskip 1.35cm
\begin{center}
{A Realistic Extension of Gauge-Mediated SUSY-Breaking Model\\
with Superconformal Hidden Sector
}
\vskip 1.2cm
Masaki Asano$^{(a)}$, Junji Hisano$^{(a,b)}$, Takashi Okada$^{(a)}$, and 
Shohei Sugiyama$^{(a)}$
\vskip 0.4cm

{$(a)$ \em Institute for Cosmic Ray Research (ICRR), \\
University of Tokyo, Kashiwa, Chiba 277-8582, Japan}\\
{$(b)$ \em Institute for the Physics and Mathematics of the Universe (IPMU) \\
University of Tokyo, Kashiwa, Chiba 277-8568, Japan}\\

\vskip 1.5cm

\abstract{
  The sequestering of supersymmetry (SUSY) breaking parameters, which is
  induced by superconformal hidden sector, is one of the solutions for
  the $\mu/B_\mu$ problem in gauge-mediated SUSY-breaking
  scenario.  However, it is found that the minimal messenger model does not derive
  the correct electroweak symmetry breaking. In this paper we 
  present a model which has the coupling of the messengers
  with the SO(10) GUT-symmetry breaking Higgs fields.  The model is
  one of the realistic extensions of the gauge mediation model with
  superconformal hidden sector. It is shown that the extension
  is applicable for a broad range of conformality breaking scale.

}
\end{center}
\end{titlepage}

\section{Introduction}

Low-energy supersymmetry (SUSY) is a very attractive model of physics
beyond the standard model (SM). In the minimal supersymmetric standard
model (MSSM), however, general SUSY-breaking masses of squarks and
sleptons induce too large FCNC and/or CP violation effects in
low-energy observables. These SUSY FCNC and CP problems should be
solved in realistic SUSY-breaking models.

Gauge-mediated SUSY breaking (GMSB) {
\cite{Dine:1981za,Dine:1993yw,Dine:1994vc,Dine:1995ag,Dine:1996xk}  }
is one of the
promising mechanisms to describe the SUSY-breaking sector in the MSSM.
The SUSY breaking is transmitted to the MSSM sector through the gauge
interaction, which induces the flavor-blind SUSY-breaking masses of
squarks and sleptons. The gaugino masses $M_a~(a=1-3)$ are generated
at one-loop level as $M_a\simeq \alpha_a/(4\pi) F_S/M_m$, and the
sfermion mass squareds are induced by two-loop diagrams so that the
sfermion masses are comparable to those for the gauginos. Here, $M_m$
and $F_S$ are the mass of the messenger and $F$-component vacuum
expectation value (VEV) of the singlet superfield $S$ in the hidden
sector, respectively, and $F_S/M_m$ is $\simeq 10$-$100$~TeV.

One of the difficulties in the model building of GMSB is the origin of the
$B_\mu$ term, which is the SUSY-breaking term corresponding to the
supersymmetric mass of the MSSM Higgs doublets, $\mu$.  $B_\mu$ has
the mass dimension two.  From viewpoints of naturalness and
electroweak symmetry breaking, $B_\mu$ and $\mu$ are required to be
comparable to the other SUSY-breaking mass parameters in the MSSM.
The correct size of $\mu$ is realized when $\mu$ is generated at
one-loop level or even when the MSSM Higgs doublets are directly
coupled with $S$ in the superpotential with a small coupling ($\sim
10^{-(2-3)}$). However, if $B_\mu$ is simultaneously induced with $\mu
$, $B_\mu$ is relatively enhanced by a one-loop factor. This problem
is sometimes called as the $\mu/B_\mu$ problem. Several mechanisms are
proposed for this problem
\cite{Dvali:1996cu,mubmuproblem,Roy:2007nz,Murayama:2007ge}.

It is pointed out in Refs.~\cite{Roy:2007nz,Murayama:2007ge} that the
$\mu/B_\mu$ problem is solved in the GMSB models with the
superconformal hidden sector (SCHS). The conformal sequestering
suppresses $B_\mu$, in addition to sfermion mass squareds $m_{\tilde f}^2
~(f=q,u,d,l,e)$
\cite{Schmaltz:2006qs}, relative to the $A$ parameters and gaugino
masses. The SUSY-breaking parameters at the scale $M_X$ at which the
conformality is broken are given as \cite{Murayama:2007ge}
\begin{eqnarray}
&{m_{\tilde f}^2=0,~(f=q,u,d,l,e)},& \nonumber\\
&{m_{H_u}^2= m_{H_d}^2=-\mu^2,~B_{\mu}=0,}&\nonumber\\
&A_u =y_u A_{H_u},~
A_d =y_d A_{H_d},~
A_l =y_l A_{H_d},
\label{bccft}
\end{eqnarray}
where $M_a~(a=1-3)\simeq \mu\simeq A_{H_u}\simeq A_{H_d}.$
Here, $m_{H_u}^2$ and $m_{H_d}^2$ are the SUSY-breaking mass squareds
for the Higgs doublets, $y_{u/d/l}$ are the Yukawa couplings for
up and down quarks and leptons, and $A_{u/d/l}$ are the $A$ parameters
for them. In addition to Eq.~(\ref{bccft}), a relationship 
$|A_{H_u} A_{H_d}| = |\mu|^2$ is also valid when the messenger sector is 
minimal. Though other arbitrary messenger sectors relax this 
relationship, it brings new sources of CP violation.

In this paper we discuss the electroweak symmetry breaking under the
boundary condition for the SUSY-breaking parameters given in
Eq.~(\ref{bccft}). It is found that the electroweak symmetry breaking
conditions have no physical solution when the messenger sector is
minimal and the GUT relation among the gaugino masses is imposed.  
We propose an extension of the minimal messenger model in which 
the messenger multiplets are coupled with the GUT-symmetry breaking 
sector in order to avoid the introducing CP violation.

It is shown that this extension makes the model phenomenologically viable
and  that it is applied for arbitrary scale for $M_X$.

The organization of the paper is as follows. In Section 2, we review
the GMSB models with SCHS. We show that the minimal messenger model has
no realistic vacuum with the electroweak symmetry broken.  In Section 3,
we propose an extension of the minimal model, which the messenger
sector is coupled with the GUT-symmetry breaking Higgs VEV. Section 4 is
devoted to conclusion.

\section{GMSB Models with SCHS and Minimal Messenger Model}

The gauge-mediated SUSY-breaking model with superconformal hidden
sector has non-trivial prediction for the SUSY-breaking parameters at
MSSM as in Eq.~(\ref{bccft}). Here, we review the derivation.  See
Ref.~\cite{Murayama:2007ge} for the detail.

We first discuss the gaugino and sfermion masses in the model as
warming up. After decoupling of the messenger multiplets with the SM
gauge quantum numbers, the following effective interactions for the
gauge and matter multiplets in the MSSM with a singlet in the hidden
sector $S$ are generated,
\begin{equation}
  {\cal L}_{eff}=
  \left\{\int d^2\theta 
  {  \sum_{a=1-3}    }
\frac12 c^a_{\lambda} \frac{S}{M_m}{\cal W}^{a\alpha}{\cal W}^{a}_{\alpha} +h.c.\right\}-
  \int d^4\theta \sum_f c^f_{m^2} \frac{S^\dagger S}{M_m^2} \tilde{f}^\dagger \tilde{f}.
\end{equation}
When $S$ gets the $F$-term VEV, $\langle S \rangle \bigr |_{\theta^2}
=F_S$, the first and second terms generate the gaugino and sfermion
masses, respectively. The coefficients $c^a_{\lambda}$ are at one-loop
level while $c^f_{m^2}$ are at two-loop level. The explicit forms for
them can be read off from formulae given in Ref.~\cite{Martin:1996zb}.

After the hidden sector enters into conformal regime at
$\Lambda_\star$, above two terms receive huge radiative correction. At
$\mu_R(<\Lambda_\star)$, the effective interactions are given as
\begin{equation}
{\cal L}_{eff}=
\left\{\int d^2\theta \sum_{a=1-3} \frac12c^a_{\lambda} Z_S^{-1/2} \frac{S}{M_m}{\cal W}^{a\alpha}{\cal W}^{a}_{\alpha} +h.c.\right\}-
\int d^4\theta \sum_f c^f_{m^2}
Z_S^{-1}  Z_{|S|^2}
 \frac{S^\dagger S}{M_m^2} \tilde{f}^\dagger \tilde{f} ,
\end{equation}
where
\begin{eqnarray}
Z_S(\mu_R)&=&\left(\frac{\Lambda_\star}{\mu_R}\right)^{3R(S)-2} 
,\nonumber\\
Z_{|S|^2}(\mu_R)&=&\left(\frac{\Lambda_\star}{\mu_R}\right)^{-\alpha_S} .
\end{eqnarray}
Here, $Z_S$ is the wave function renormalization of $S$ and $R(S)$ is
the $R$ charge for $S$. When $S$ is singlet under the hidden gauge
groups, $R(S)$ is larger than $2/3$ so that $Z_S(\mu_R)>1$. The 1PI
contribution to operators $S^\dagger S$ is parametrized by $\alpha_S$
in the above equation. 

The gaugino masses $M_a$ at $M_X$, at which the conformality is
broken, are given as
\begin{eqnarray}
  M_a&=& c^a_{\lambda} Z_S^{-1/2}(M_X) \frac{F_S}{M_m}.
\end{eqnarray}
When $\alpha_S>0$, the sfermion masses are suppressed, and the
conformal sequestering is realized as \cite{Schmaltz:2006qs}
\begin{eqnarray}
{m_{\tilde f}^2=0,~(f=q,u,d,l,e)}.
\end{eqnarray}

Next, let us move to the Higgs sector. Here, the messenger sector is
assumed to be minimal among models where the $\mu$ term is generated
by one-loop diagrams. Then, the messenger multiplets are embedded in
SU(5) $\bf 10$ and $\bf 10^\star$-dimensional multiplets.\footnote{
  Even when the messengers are SU(5) $\bf 5$ and $\bf
  5^\star$-dimensional multiplets, the $\mu$ term is generated if additional
  SU(5) singlets are also introduced. However, when the singlets are
  coupled with $S$, the arbitrary phases in the interactions generate CP-violating phases in
 the $A$ and $B_\mu$ terms.
}
The
messenger multiplets have an interaction with the Higgs doublets $H_u$
and $H_d$ in the superpotential,
\begin{eqnarray}
W= \lambda_u H_u Q_m U_m+ \lambda_d H_d \Bar{Q}_m \Bar{U}_m 
+(\kappa S + M_m)(Q_m \Bar{Q}_m + U_m \Bar{U}_m+E_m \Bar{E}_m),
\end{eqnarray}
where $Q_m$, $U_m$, and $E_m$ ($\Bar{Q}_{m}$, $\Bar{U}_{m}$,
and $\Bar{E}_{m}$), which come from the SU(5) $\bf 10$ ($\bf
10^\star$)  multiplet, have SU(5) symmetric mass and
interaction terms. 

Integration of the messenger sector leads to the effective interactions
of the Higgs doublets with $S$ as
\begin{eqnarray}
{\cal L}_{eff}&=&
-\int d^4\theta \left\{
 c_{\mu} \frac{S^\dagger}{M_m} H_d H_u
+c_{B_\mu} \frac{S^\dagger S}{M_m^2} H_d H_u
\right.
\nonumber\\
&&\left.
+ c_{A_u} \frac{S}{M_m} H_u^\dagger H_u
+ c_{A_d} \frac{S}{M_m} H_d^\dagger H_d+h.c.\right\}
\nonumber\\
&&
-\int d^4\theta \left\{
c^{H_u}_{m^2} \frac{S^\dagger S}{M_m^2} H_u^\dagger H_u
+c^{H_d}_{m^2} \frac{S^\dagger S}{M_m^2} H_d^\dagger H_d\right\}.
\label{higgseff}
\end{eqnarray}
Here, the coefficients of the operators, $c_{\mu}$,
$c_{B_\mu}$, ${c_{A_u} }$, and $c_{A_d}$ are generated at one-loop level,
\begin{eqnarray}
c_{\mu}=-3\frac{\lambda_u\lambda_d}{(4\pi)^2} \kappa^*,&&
c_{B_\mu}=-3 \frac{\lambda_u\lambda_d}{(4\pi)^2} |\kappa|^2,\nonumber\\
c_{A_u}=+3 \frac{|\lambda_u|^2}{(4\pi)^2} \kappa,&&
c_{A_d}=+3 \frac{|\lambda_d|^2}{(4\pi)^2} \kappa,
\end{eqnarray}
while $c^{H_u}_{m^2}$ and $c^{H_d}_{m^2}$ are vanishing
at one-loop level. 

After the hidden sector enters into conformal regime, the effective
interactions become 
\begin{eqnarray}
{\cal L}_{eff}&=&
-\int d^4\theta \left\{
 c_{\mu} Z_S^{-1/2}\frac{S^\dagger}{M_m} H_d H_u
\right.
\nonumber\\
&&
\left.
+ Z_S^{-1}  \left[Z_{|S|^2} c_{B_\mu} 
+(Z_{|S|^2}-1)(c_{\mu} c_{A_u}+c_{\mu} c_{A_d})\right]
\frac{S^\dagger S}{M_m^2} H_d H_u
+h.c.\right\}
\nonumber\\
&&
-\int d^4\theta \left\{
c_{A_u} Z_S^{-1/2} \frac{S}{M_m} H_u^\dagger H_u
+h.c.
\right.
\nonumber\\
&&
\left.
+ Z_S^{-1}\left[
Z_{|S|^2}c^{H_u}_{m^2}
+(Z_{|S|^2}-1)
(|c_{A_u}|^2+|c_{\mu}|^2)
\right]
\frac{S^\dagger S}{M_m^2} H_u^\dagger H_u
{+(H_u\leftrightarrow H_d) } \right\} 
.
\end{eqnarray}
The terms proportional to $(Z_{|S|^2}-1)$  come from diagrams with the
Higgs doublet exchange. Since the tree-level diagrams with the Higgs
exchange do not contribute to the effective Lagrangian, one is
subtracted from $Z_{|S|^2}$ there. Therefore, the SUSY-breaking terms
in the Higgs sector at $M_X$ are
\begin{eqnarray}
&{m_{H_u}^2= m_{H_d}^2=-\mu^2,~B_{\mu}=0,}&\nonumber\\
&A_u =y_u A_{H_u},~
A_d =y_d A_{H_d},~
A_l =y_l A_{H_d},&
\label{higgsbc}
\end{eqnarray}
 when
$Z_{|S|^2}(M_X)\ll1$. Here, 
\begin{eqnarray}
\mu&=&  c_{\mu} Z_S^{-1/2}\frac{F_S^\dagger}{M_m}
=
-3\frac{\lambda_u\lambda_d}{(4\pi)^2} 
Z_S^{-1/2}\frac{\kappa^*F_S^\dagger}{M_m}, \nonumber\\
A_{H_u/H_d}&=& - c_{A_{u/d}} Z_S^{-1/2} \frac{F_S}{M_m}
=
-3 \frac{|\lambda_{u/d}|^2}{(4\pi)^2}Z_S^{-1/2} \frac{\kappa F_S}{M_m}.
\label{muaud}
\end{eqnarray} 
In the derivation of Eq.~(\ref{higgsbc}), we redefined the Higgs
doublets as  $H_{u/d} - c_{A_{u/d}} Z_S^{-1/2} \frac{S}{M_m}
H_{u/d}\rightarrow H_{u/d}$ .  Since we now derived the 
SUSY-breaking terms in the minimal messenger model, Eq.~(\ref{muaud}) 
satisfies the relationship $|A_{H_u} A_{H_d}| = |\mu |^2$.
In Appendix we give formulae for the
SUSY-breaking terms of the Higgs sector in more general messenger cases.

Now we discuss the electroweak symmetry breaking in the GMSB models with
SCHS. The minimization condition of the Higgs potential at tree
level results in
\begin{eqnarray}
\sin2\beta&=& -\frac{2B_\mu}{m_1^2+m_2^2},
\label{ewsb1}\\
m_Z^2&=&-\frac{m_{1}^2-m_{2}^2}{\cos 2 \beta}
-(m_{1}^2+m_{2}^2),
\label{ewsb2}
\end{eqnarray}
where $m_1^2\equiv(m_{H_d}^2+\mu^2)$ and
$m_2^2\equiv(m_{H_u}^2+\mu^2)$. In the GMSB models with SCHS, the Higgs boson
mass squareds are zero at tree level even after including the
supersymmetric mass $\mu$.  Thus, the electroweak symmetry breaking
and the stability of the Higgs boson potential are sensitive to the
radiative corrections to them.

\begin{figure}[t]
\begin{center}
\begin{tabular}{cc}
\includegraphics[scale=0.6, angle = 0]{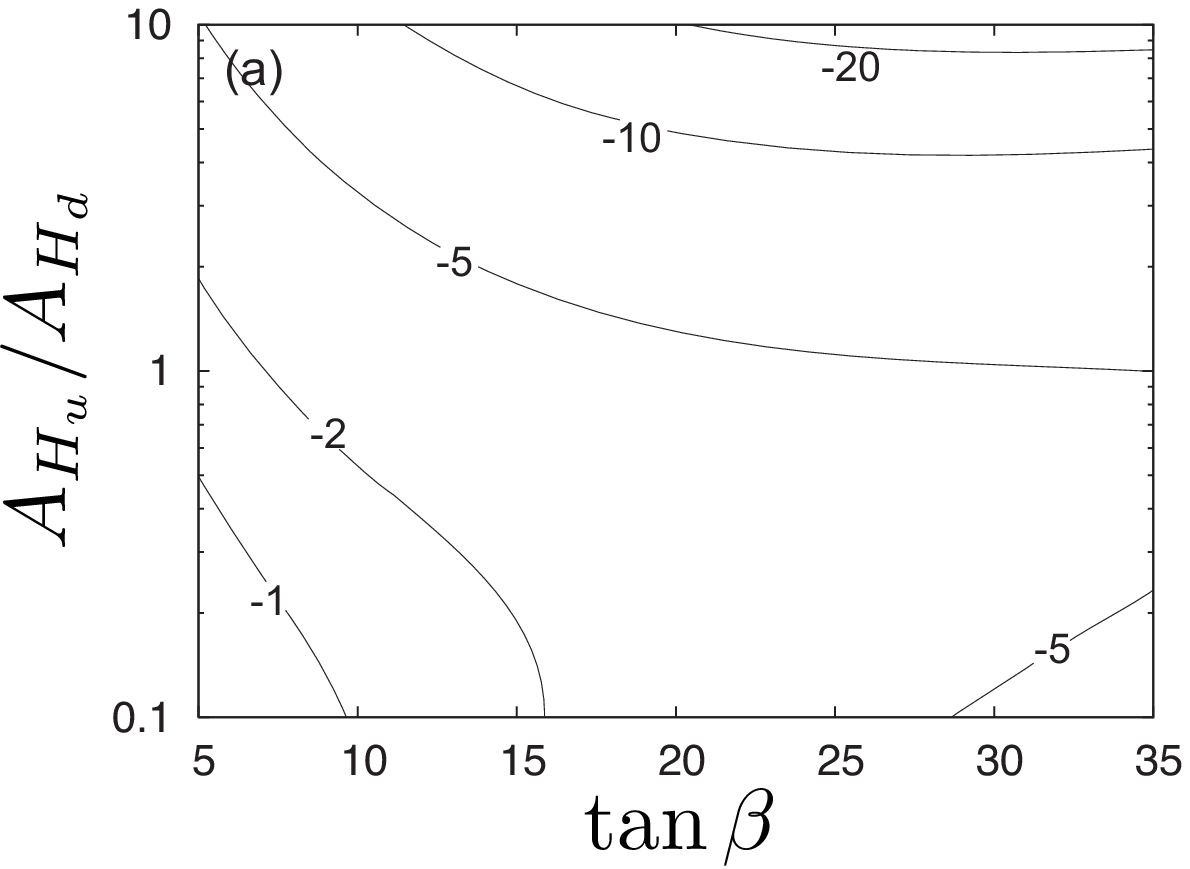}&
\includegraphics[scale=0.6, angle = 0]{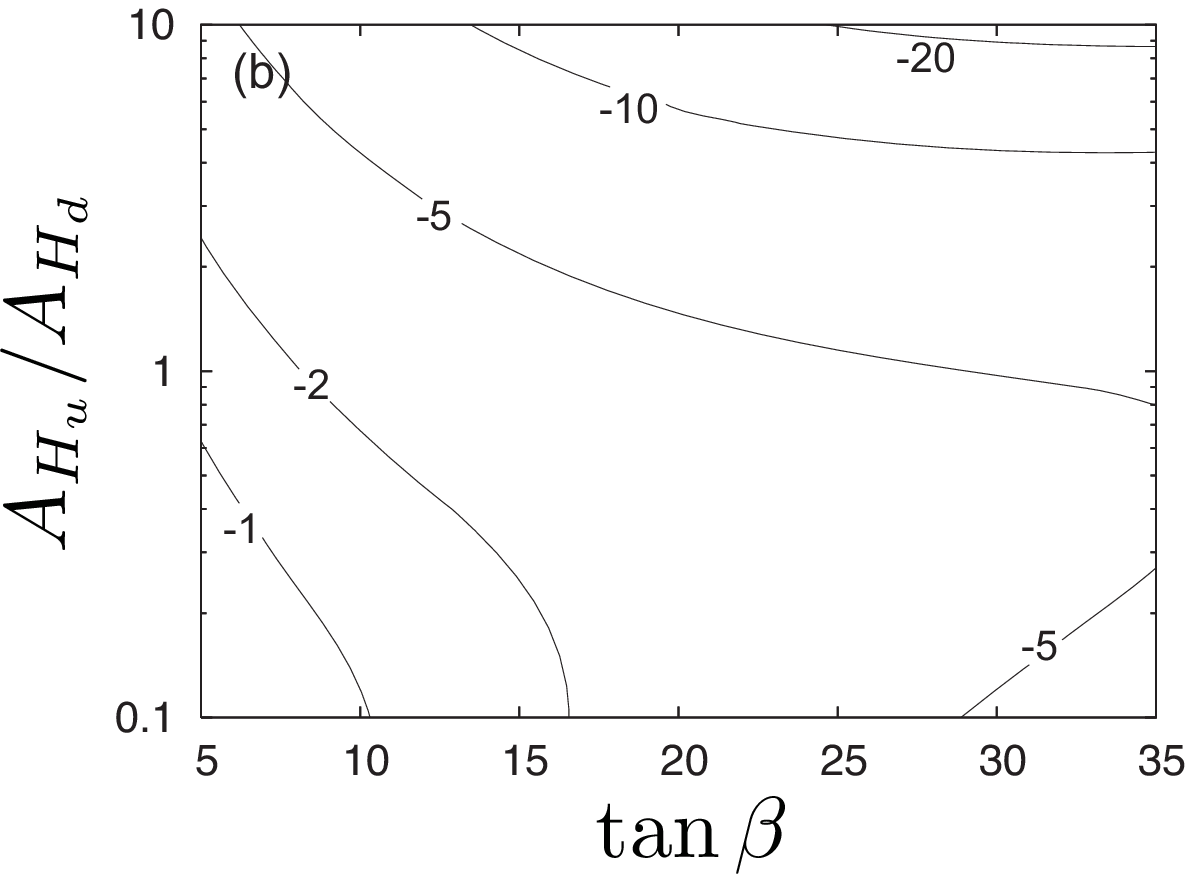}
\end{tabular}\\
\includegraphics[scale=0.6, angle = 0]{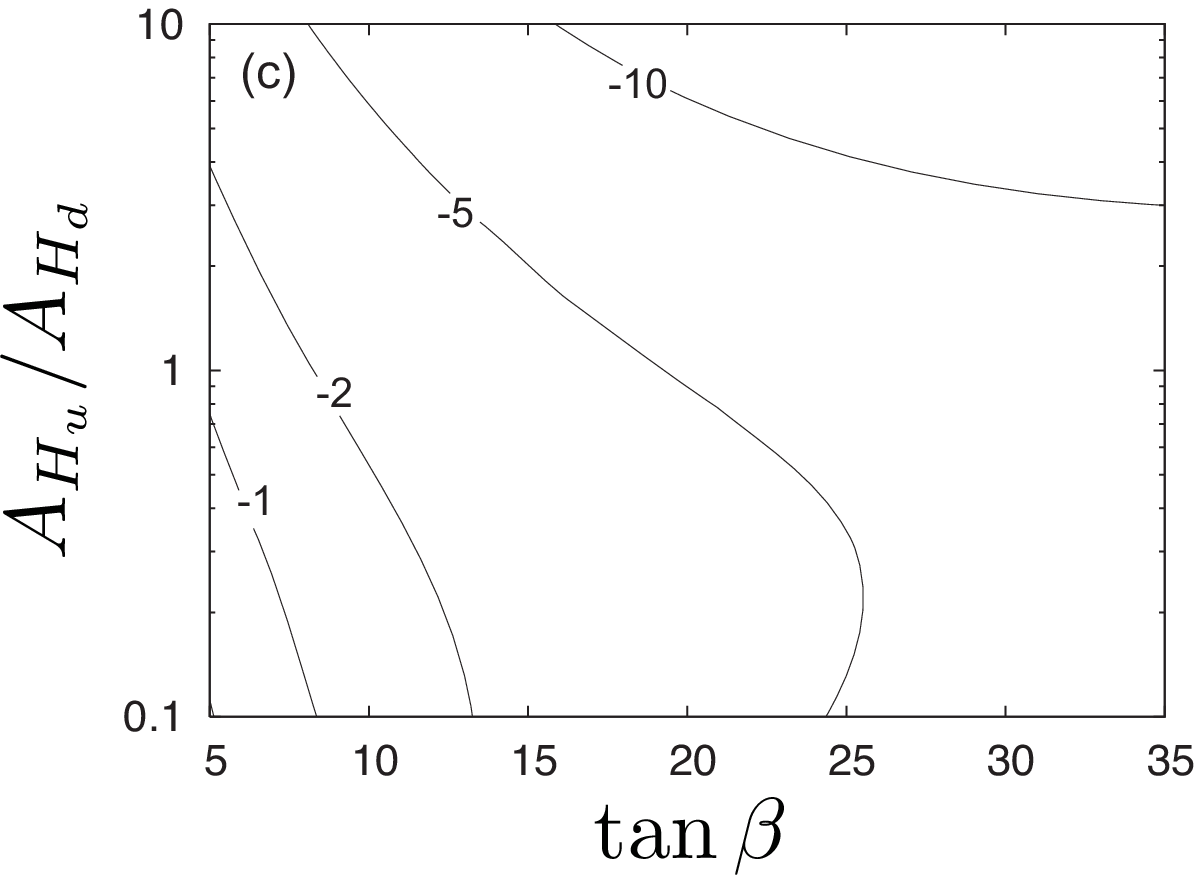}
\caption{\label{fig:minimal} 
$m_A^2/\mu^2$ as a function of $A_{H_u}/A_{H_d}$ and $\tan\beta$.
We take $M_X=10^{14}$~GeV in (a), $M_X=10^{10}$~GeV in (b), and 
$M_X=10^{6}$~GeV in (c). 
}
\end{center}
\end{figure}

In Fig.~\ref{fig:minimal} we show the pseudo-scalar Higgs mass squared
$m_A^2(\equiv m_1^2+m_2^2)$ normalized by $\mu^2$ as a function of
$A_{H_u}/A_{H_d}$ and $\tan\beta$ in the minimal messenger model. We take
$M_X=10^{14}$~GeV in (a), $M_X=10^{10}$~GeV in (b), and
$M_X=10^{6}$~GeV in (c), and $m_A^2/\mu^2$ is evaluated at
$m_{SUSY}=1$~TeV. Here, the messengers have SU(5) symmetric mass terms
so that the gaugino masses obey the GUT relation.  In the minimal
model, the input parameters are the gluino mass $M_{3}$,
$\mu$ and $A_{H_u}/A_{H_d}$ in addition to $M_X$, two of which are fixed
by two Higgs VEVs. Eq.~(\ref{ewsb1}) determines the ratio of $\mu$ and
$M_{3}$, and then $m_A^2/\mu^2$.

It is found from Fig.~\ref{fig:minimal} that $m_A^2$ is always
negative. This implies that the vacuum is not stabilized. In order to
qualitatively understand this result, we derived approximation of the
mass parameters in the Higgs potential from the renormalization-group equations (RGE)  as
\begin{eqnarray}
m_1^2\equiv(m_{H_d}^2+\mu^2)(m_{SUSY})
&=&
(3\alpha_2+\alpha_Y)\mu^2 t_{SUSY}
+(3\alpha_2 M_2^2+\alpha_Y M_1^2) t_{SUSY}
\nonumber\\
&&-3\alpha_t\mu^2 t_{SUSY} ,
\nonumber\\
m_2^2\equiv(m_{H_u}^2+\mu^2)(m_{SUSY})
&=&
(3\alpha_2+\alpha_Y)\mu^2 t_{SUSY}
+(3\alpha_2 M_2^2+\alpha_Y M_1^2) t_{SUSY}
\nonumber\\
&&
-3\alpha_t A_{H_u}^2 t_{SUSY}
-16\alpha_3 \alpha_t (M_3^2+A_{H_u} M_3) t_{SUSY}^2 , 
\nonumber\\
B_\mu/\mu(m_{SUSY})
&=&
(3\alpha_2 M_2+\alpha_Y M_1-3\alpha_t A_{H_u}) t_{SUSY}
\nonumber\\
&&
-8\alpha_t \alpha_3 M_3 t_{SUSY}^2 ,
\label{log}
\end{eqnarray}
where $t_{SUSY}=\log(M_X/m_{SUSY})/2\pi$, $\alpha_t(\equiv
y_t^2/4\pi)$ and $\alpha_a$ $(a=Y,2,3)$ are for the top-quark Yukawa
and gauge coupling constants, respectively. Here, we include the
one-loop contributions due to the electroweak and top-quark Yukawa
interactions and two-loop contributions due to the strong one. The
later one is comparable to the one-loop terms when the gluino mass is
larger than others, as in the GUT relation. These equations are
  semi-quantitatively valid when $\alpha_{a}t_{SUSY},~\alpha_{t}
  t_{SUSY} \ll 1$. Even when $\alpha_{a} t_{SUSY}~, \alpha_{t}
  t_{SUSY} \sim O(1)$, we can guess the qualitative behaviors, such as
  relative signs and sizes among the terms, using the equations.

It is found that $A_{H_u/H_d}$ are negative in Eq.~(\ref{muaud}).
This implies that the one-loop contributions to $B_\mu/\mu$ are
constructive. Sizable values of $B_\mu/\mu$ lead to suppression of
$\mu/M_{3}$ from Eq.~(\ref{ewsb1}) for $\tan\beta\gsim 1$. In
those cases the two-loop contribution, which enhanced by the gluino mass,
derives $m_A^2$ to be negative. When $\tan\beta\simeq 1$,
$\mu/M_{3}\sim 1$ is possible. However, it is found from the
figure that $m_A^2$ is still negative.

One of the solutions for the problem is introduction of non-zero
$B_\mu$ at $M_X$. If $Z_{|S|^2}(M_X)$ is {\it accidentally} around 
$O(10^{-(2-3)})$, $B_\mu$ keeps its sizable value at $M_X$. However, its
sign is positive relatively to $\mu$ since 
\begin{eqnarray}
B_\mu &=&-3 \frac{\lambda_u\lambda_d}{(4\pi)^2} 
Z_S^{-1}Z_{|S|^2}\frac{|\kappa|^2|F_S|^2}{M_m^2} .
\end{eqnarray}
This is constructive to the RGE contribution to $B_\mu$, while the
deconstructive interference is rather required for the electroweak
symmetry breaking. This is also noticed in Ref.~\cite{Cho:2008fr}.
If the operator $S^\dagger S$ is  mixed with other operators whose
$D$-component VEVs are non-vanishing, the sign of the contribution 
to $B_\mu$ may be changed.

The second solution is extension of the messenger sector.  When
introducing multiple messengers with different supersymmetric masses
and couplings with $S$, the deconstructive interference in $B_\mu$ is
possible.  However, arbitrary introduction of the messengers leads to
CP phases in the SUSY-breaking parameters. That is not favored from 
phenomenological viewpoints.

\section{Extension}

One of the extensions of the GMSB with SCHS without introducing CP
violation is introduction of coupling of the messengers with the 
GUT-symmetry breaking Higgs fields. Let us consider following superpotential;
\begin{eqnarray}
W&=&\lambda_u H_u \psi \psi+ \lambda_d H_d \Bar{\psi} \Bar{\psi} 
+(\kappa S + \zeta \Sigma) \bar{\psi} \psi.
\end{eqnarray}
Here, $\psi$ and $\Bar{\psi}$ are the messengers and $\Sigma$ is the
GUT-symmetry breaking Higgs fields. The messengers are $\bf 10$ and
$\bf 10^\star$-dimensional multiplets in the SU(5) GUTs, and $\bf 16$ and
$\bf 16^\star$ in the SO(10) GUTs.

It is found that this extension does not work well in the SU(5) GUTs.
When the SU(5) breaking Higgs field is a ${\bf 24}$-dimensional multiplet, the
messenger masses are proportional to their hypercharges so that the
bino mass is zero at one-loop level.  When the SU(5) breaking Higgs
field is a ${\bf 75}$-dimensional multiplet, the SU(2) doublet messenger
quark and singlet messenger quark masses are degenerate with the opposite
sign. Then, $\mu$ and $A$ parameters are zero at one-loop level.

These problems are resolved when the messenger masses are generated by
the higher-dimensional operators with $\Sigma$. In those cases
the colored messengers are relatively lighter so that the gluino 
becomes heavier. From the electroweak symmetry breaking condition
in Eq.~(\ref{ewsb2}), which is reduced $m_Z^2\simeq -2 m_{2}^2$
for $\tan\beta\gsim 1$,  larger $M_{3}$ leads to larger $\mu$.
However,  $m_A^2(\simeq m_1^2-1/2 m_Z^2)$  is likely to be
tachyonic due to large $\mu$.

Thus, we consider the SO(10) GUTs. Here, we assume that $\Sigma$ is a
${\bf 45}$-dimensional multiplet. 
The messenger masses are
given by hypercharge $Q_Y$ and $(B-L)$ charge of the messengers $Q_{B-L}$
\footnote{
In this paper,  the assignment of ${(B-L)}$ charges for quarks and leptons
are $1/3$ and $(-1)$, respectively.
}, because
\begin{eqnarray}
\zeta\langle  \Sigma \rangle &=& Q_Y M_Y + Q_{B-L} M_{B-L} .
\end{eqnarray}
%

In the following, we consider a case where only the SU(5) $\bf 10$ and
$\bf 10^\star$-dimensional components of the ${\bf 16}$ and $\bf
16^\star$-dimensional multiplets become effective in generation of
the SUSY-breaking terms in the MSSM.
 This is only for simplicity,
because when the SO(10) full multiplets contribute to SUSY-breaking
mediation, the $M_Y/M_{B-L}$ dependence of soft breaking parameters 
is more complicated.
Actually, it is realized when an SO(10) ${\bf
  10}$-dimensional multiplet is introduced in the messenger sector.
In that case, we can add following terms to the superpotential,
\begin{eqnarray}
W&=&
  f_u \psi \phi \psi_H + f_d \Bar{\psi} \phi \Bar{\psi}_H 
+ \frac{1}{2} M \phi \phi ,
\end{eqnarray}
where $\psi_H$($\Bar{\psi}_H$) are ${\bf 16}$(${\bf 16^\star}$)-dimensional multiplets and $\phi$ is a
${\bf 10}$-dimensional matter multiplet. 
The SU(5) $\bf 5$ and $\bf 5^\star$ multiplets of ${\bf 16}$ 
and $\bf 16^\star$ are decoupled when SU(5) singlets of 
$\psi_H$ and $\Bar{\psi}_H$ have non-zero vacuum expectation values. 
%
%

\begin{figure}[t]
\begin{center}
\begin{tabular}{cc}
\includegraphics[scale=0.6, angle = 0]{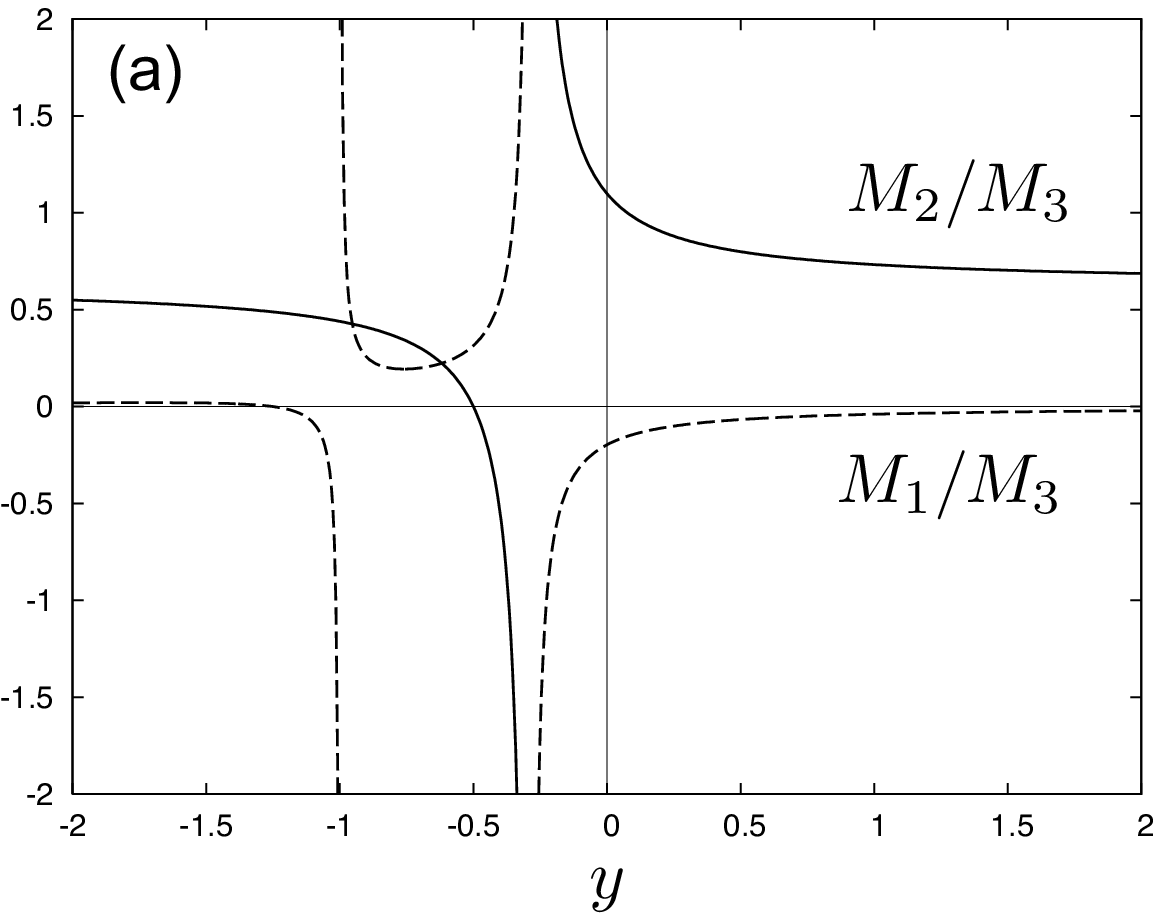}&
\includegraphics[scale=0.6, angle = 0]{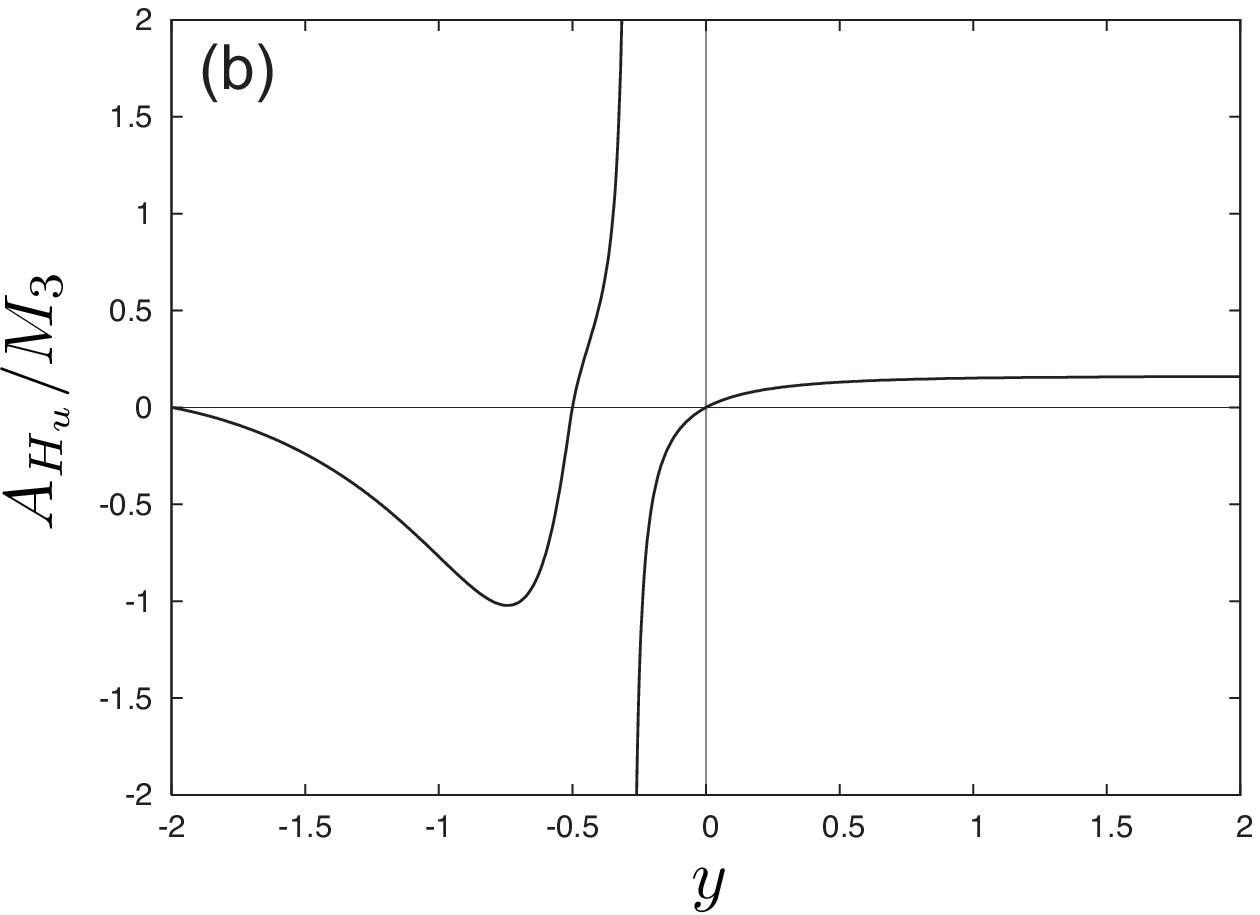}
\end{tabular}\\
\caption{\label{fig:ratio} 
Ratios $M_{2}/M_{3}$ and  $M_{1}/M_{3}$ at 1~TeV (a) 
and  $A_{H_u}/M_{3}$ at $M_X$ (b)
as functions of $y(\equiv M_Y/M_{B-L})$. 
Here, $\lambda_u=g_3$ for simplicity.
}
\end{center}
\end{figure}

In Fig.~\ref{fig:ratio} ratios $M_{2}/M_{3}$ and $M_{1}/M_{3}$
at 1~TeV and $A_{H_u}/M_{3}$ at $M_X$ are shown as functions of
$y(\equiv M_Y/M_{B-L})$. Notice that when $y>0$, $A_{H_u}/M_{3}$ is 
positive and gluino is lighter than twice the mass of wino.
These are welcome to the
electroweak symmetry breaking as discussed above. In fact, we could
easily find the solutions which are phenomenologically viable.  We
studied the other regions. However, though we found points to be 
consistent with the electroweak symmetry breaking conditions, 
their spectrums are quite light so that they are experimentally
excluded.

We show mass spectra and 
the branching ratio BR($b \rightarrow s \gamma$) at several points, 
$M_X = 10^8, 10^{11},$ and $10^{14}$~GeV, in Table~\ref{mass_spectrum} using 
SuSpect 2.41 \cite{Djouadi:2002ze} and SusyBSG 1.1.2 
\cite{Degrassi:2007kj}. All of them are consistent with the Higgs boson 
mass bound, sparticle mass bounds ~\cite{Amsler:2008zz} and branching 
ratio of $b \rightarrow s \gamma$~\cite{Barberio:2007cr};
\begin{eqnarray}
{\rm Br} ( b \to s \gamma) &=& 
       (355 \pm 24^{+9}_{-10} \pm3) \times 10^{-6} .
\label{bound_bsg}
\end{eqnarray}

In all sample points, the right-handed slepton masses are very small
compared with other sparticle masses.
As we have seen in Eq.~(\ref{bccft}), the scalar fermion soft masses are 
nearly zero at $M_X$. In addition, when $y>0$, the bino is light 
compared to the wino and gluino. As a result, the right-handed slepton 
masses are such small in this model. 

Using SuSpect 2.34, we also calculated supersymmetric contributions to 
the anomalous magnetic moment of the muon, $a_{\mu} = (g-2)_{\mu}/2$. 
The comparison between the measurements~\cite{Bennett:2006fi} and the 
SM theoretical predictions~\cite{Hagiwara:2006jt} for $a_{\mu}$ is 
\begin{eqnarray}
\Delta a_{\mu} &=& a^{\rm exp}_{\mu} - a^{\rm SM}_{\mu}
                = (30.2 \pm 8.7)\times 10^{-10}.
\end{eqnarray}
The left-handed sleptons are so heavy that the SUSY contribution to
$a_{\mu}$ is suppressed. When the deviation is confirmed in future,
this model would be disfavored.

%
\begin{table}[p]
\centering
\begin{tabular}{|c|c|c|c|c|c|}
\hline
$\tan \beta$                    
                            & 10
                            & 10
                            & 10 \\
%
$y$
                            & 0.281
                            & 0.150
                            & 0.0452 \\
$A_{H_u}/A_{H_d}$                  
                            & 11.5
                            & 7.16
                            & 1.00 \\
$M_X$                  
                            & $10^8$ [GeV]
                            & $10^{11}$ [GeV]
                            & $10^{14}$ [GeV] \\
\hline \hline
$\widetilde{g}$          
                         & 2906
                         & 1553
                         & 1304 \\
\hline
$\widetilde{\chi}^{\pm}_1$ 
                         & 1049
                         & 918.9
                         & 354.8 \\
$\widetilde{\chi}^{\pm}_2$ 
                         & 2637
                         & 1443
                         & 1303 \\
\hline
$\widetilde{\chi}^0_1$     
                         &  293.3
                         &  191.8
                         &  203.7 \\
$\widetilde{\chi}^0_2$     
                         &  1049
                         &  918.1
                         &  354.0 \\
$\widetilde{\chi}^0_3$     
                         &  1052
                         &  925.3
                         &  365.1 \\
$\widetilde{\chi}^0_4$     
                         &  2637
                         &  1443
                         &  1303 \\
\hline
$\widetilde{t}_1$    
                         & 1777
                         & 1013
                         & 758.4 \\
$\widetilde{t}_2$    
                         & 2255
                         & 1424
                         & 1319 \\
${(\widetilde{u},\widetilde{c})}_{L,R}$  
                         & 2330, 2030   
                         & 1477, 1223
                         & 1403, 1077    \\
\hline
$\widetilde{b}_1$    
                         & 2024   
                         & 1218   
                         & 1066    \\
$\widetilde{b}_2$    
                         & 2238 
                         & 1403   
                         & 1303    \\
${(\widetilde{d},\widetilde{s})}_{L,R}$  
                         & 2331, 2029
                         & 1479, 1223
                         & 1405, 1075 \\
\hline
$\widetilde{\tau}_1$  
                          & 98.41
                          & 101.3
                          & 128.5 \\
$\widetilde{\tau}_2$  
                          & 1132
                          & 825.7
                          & 904.6 \\
${(\widetilde{e},\widetilde{\mu})}_{L,R}$ 
                          & 1132, 104.6
                          & 825.5, 102.7
                          & 906.1, 148.0\\
\hline
$\widetilde{\nu}_{\tau}$  
                          & 1129
                          & 821.9
                          & 901.3   \\
$\widetilde{\nu}_{e,\mu}$ 
                          & 1129
                          & 821.9
                          & 902.7   \\
\hline
$h$                    & 117.8
                         & 115.4
                         & 114.5 \\
$H$             
                         & 1142
                         & 810.2
                         & 903.0 \\
$A$              
                         & 1142
                         & 809.9
                         & 902.8 \\
$H^{\pm}$            
                         & 1145
                         & 814.1
                         & 906.7 \\
\hline \hline
BR($b\rightarrow s \gamma$)            
                         & 3.41$\times 10^{-4}$
                         & 3.68$\times 10^{-4}$
                         & 3.77$\times 10^{-4}$ \\
\hline
$\Delta a_{\mu}$            
                         & -1.17$\times 10^{-10}$
                         & -1.29$\times 10^{-10}$
                         & -4.86$\times 10^{-10}$ \\
\hline
\end{tabular}
\caption{
Sparticle and Higgs boson mass spectra (in units of GeV) 
in $m_t=171.2 ~{\rm GeV}$.}
\label{mass_spectrum}
\end{table}

\section{Conclusion}


We have studied the electroweak symmetry breaking for the 
GMSB models with SCHS which solve the $\mu/B_\mu$ problem
by the conformal sequestering. 
It is found that the correct electroweak symmetry breaking 
is not derived in the minimal messenger model with GUT relation 
among the gaugino masses.

In this paper we also propose an extension of the minimal model
which has the coupling of the messengers with the SO(10) GUT-symmetry
breaking Higgs fields. This is one of the minimal extension
to realize a realistic model without introducing CP violation in the 
SUSY-breaking terms.
The extended messenger sector allows us to change the gaugino mass 
relation and the signs of $A$ and $B_{\mu}$ parameters. Thus, 
the model can induce the correct electroweak symmetry breaking
even if the $B_\mu$ is significantly suppressed ($B_\mu$ = 0)
at the conformality breaking scale, $M_X$. Moreover, the model can be 
applied for a broad range of the $M_X$.
In a case where all but the SU(5) $\bf 10$ and $\bf 10^\star$ multiplets
of the ${\bf 16}$ and $\bf 16^\star$ are decoupled, for example, 
we have presented mass spectra in several values of $M_X$.
They are consistent with the lightest Higgs boson mass bound, 
sparticle mass bounds, and branching ratio of $b \rightarrow s \gamma$.

\section*{Acknowledgement} This work was supported by World Premier
International Center Initiative (WPI Program), MEXT, Japan.  The work
of JH was also supported in part by the Grant-in-Aid for Science
Research, Japan Society for the Promotion of Science (No.~20244037 and
No.~2054252).

\section*{Appendix}

Here, we give formulae for the SUSY-breaking terms of the Higgs sector 
in the following messenger model,
\begin{eqnarray}
W= \lambda_u H_u \phi_1 \phi_2 + \lambda_d H_d \Bar{\phi_1} \Bar{\phi_2}  
+(\kappa_1 S + M_1)\phi_1 \Bar{\phi_1} + (\kappa_2 S + M_2)\phi_2 \Bar{\phi_2}
\end{eqnarray}
where $\phi_{1,2}$ and ${\Bar{\phi}}_{1,2}$ are the messengers, and $S$
is a singlet in the hidden sector, which acquires a non-zero
$F$-component VEV, $\langle S \rangle \bigr |_{\theta^2} =F_S$, at the
SUSY-breaking scale. After integrating out the messengers, this
SUSY-breaking VEV generates $\mu$, $B_{\mu}$, $A$ terms, and the
gaugino masses. The SUSY-breaking terms of the Higgs sector are
parametrized as 
\begin{eqnarray}
V &=& m_{H_u}^2 |H_u|^2 + m_{H_d}^2  |H_d|^2
+(B_\mu H_d H_u  + h.c.)
\nonumber\\
&&
+(      A_{H_u} f_u u  H_u  Q 
+ A_{H_d} f_d d H_d  Q
+ A_{H_d} f_e e H_d  L +h.c. ) .
\end{eqnarray}
The $\mu$, $B_\mu$, and $A$ terms are given as 
\begin{eqnarray}
\mu &=&  -\frac{\lambda_u \lambda_d}{(4 \pi)^2} \biggl[
   x_{21}\ g(x_{21})\frac{\kappa^*_1 F^*_S}{M_1}+x_{12}\ g(x_{12})\frac{\kappa^*_2 F^*_S}{M_2}\biggr],
\nonumber\\
B_{\mu} &=& - \frac{\lambda_u \lambda_d}{(4 \pi)^2}
\biggl[
 f_1(x_{21}) \frac{|\kappa_1 F_S|^2}{M_1^2}
+f_1(x_{12}) \frac{|\kappa_2 F_S|^2}{M_2^2} \nonumber\\
&&+f_2(x_{21}) \frac{\kappa_1^* \kappa_2   |F_S|^2}{M_1^2}
+f_2(x_{12}) \frac{\kappa_1   \kappa_2^* |F_S|^2}{M_2^2}
\biggr] ,\nonumber\\
A_{H_u} &=& -\frac{|\lambda_u|^2}{(4\pi)^2}\biggl[g(x_{21})\frac{\kappa_1 F_S}{M_1}+ g(x_{12})\frac{\kappa_2 F_S}{M_2}\biggr],
\nonumber\\
A_{H_d} &=& -\frac{|\lambda_d|^2}{(4\pi)^2}\biggl[g(x_{21})\frac{\kappa_1 F_S}{M_1}+ g(x_{12})\frac{\kappa_2 F_S}{M_2}\biggr],
\end{eqnarray}
where the supersymmetric messenger masses $M_1$ and $M_2$ are taken
real, and $x_{12}= M_1/M_2$. The mass functions $g(x)$, $f_1(x)$ and
$f_2(x)$ are
\begin{eqnarray}
g(x)&=& \frac{1}{(1-x^2)^2}(1-x^2+x^2 \log x^2),\nonumber\\
f_1(x)&=& \frac{1}{(1-x^2)^3}x(1-x^4+2x^2\log x^2),\nonumber\\
f_2(x)&=& - \frac{1}{(1-x^2)^3} x^2(2(1-x^2)+(1+x^2)\log x^2) .
\end{eqnarray}
These three functions are positive definite.

\end{document}